\documentclass[acmtog]{acmart}

\usepackage{booktabs} 
\usepackage{mathtools,colortbl}
\usepackage{appendix}
\usepackage{bm}
\usepackage{stmaryrd} 
\usepackage{cleveref} 

\usepackage[ruled]{algorithm2e} 

\SetAlFnt{\small}
\SetAlCapFnt{\small}
\SetAlCapNameFnt{\small}
\SetAlCapHSkip{0pt}

\acmSubmissionID{2119}
\acmJournal{TOG}

\copyrightyear{2026}
\acmYear{2026}
\setcopyright{cc}
\setcctype{by}
\acmConference[SA Conference Papers '26]{SIGGRAPH Asia 2026 Conference Papers}{December 01--04, 2026}{Kuala Lumpur, Malaysia}
\acmBooktitle{SIGGRAPH Asia 2026 Conference Papers (SA Conference Papers '26), December 01--04, 2026, Kuala Lumpur, Malaysia}
\acmDOI{10.1145/3829340.3842314}
\acmISBN{979-8-4007-2842-6/2026/12}

\newcommand{\vect}[1]{\mathbf{#1}}        
\newcommand{\xx}{\vect{x}}                
\newcommand{\pp}{\vect{p}}                
\newcommand{\dd}{\vect{d}}                
\newcommand{\diro}{\bm{\omega}}                
\newcommand{\diri}{\bm{\omega}_i}              
\newcommand{\bsdfsym}{{\rho}}             
\newcommand{\xb}{\overline{\vect{x}}}          

\newcommand{\surf}{\mathcal{S}}           
\newcommand{\manifold}{\mathcal{M}}       
\newcommand{\n}{\vect{n}}                 
\newcommand{\hemi}{\mathcal{H}}           

\newcommand{\Lo}{L_o}                     
\newcommand{\flux}{\Phi}                  
\newcommand{\appflux}{\tilde{\Phi}}       

\newcommand{\pdf}{p}                      
\newcommand{\pdfPos}{p_{\text{pos}}}      
\newcommand{\pdfDir}{p_{\text{dir}}}      

\newcommand{\netStudent}{f_{\phi}}        

\newcommand{\rev}[1]{\textcolor{black}{#1}}
\newcommand{\shorten}[1]{\textcolor{black}{#1}}

\begin{document}
\title{PureLight: Learning Complex Luminaires with Light Tracing}

\author{Pedro Figueiredo}
\orcid{0000-0002-3807-1512}
\affiliation{
\institution{Texas A\&M University}
\country{USA}
}
\email{pedrofigueiredo@tamu.edu}

\author{Zixuan Li}
\orcid{0009-0004-2424-9529}
\affiliation{
\institution{Nankai University}
\country{China}
}

\author{Beibei Wang}
\orcid{0000-0001-8943-8364}
\affiliation{
\institution{Nanjing University}
\country{China}
}

\author{Milo{\v{s}} Ha{\v{s}}an}
\orcid{0000-0003-3808-6092}
\affiliation{
  \institution{NVIDIA}
  \country{USA}
}

\author{Nima Khademi Kalantari}
\orcid{0000-0002-2588-9219}
\affiliation{
\institution{Texas A\&M University}
\country{USA}
}
\email{nimak@tamu.edu}

\renewcommand\shortauthors{Figueiredo et al.}

\begin{abstract}
We propose a neural formulation for estimating the appearance of complex luminaires. We focus on challenging luminaires with complex light transport (e.g., small emitters enclosed by multiple specular layers) that are difficult for (bidirectional) path tracing. To this end, we use light tracing to construct paths from emitters to the exit surfaces and formulate appearance estimation as a distribution learning problem. Specifically, we model the probability density function (pdf) of outgoing radiance on the exit surfaces using a large normalizing flow network, and recover the outgoing radiance as the product of the estimated pdf and flux. To enable efficient inference, we distill the learned appearance into a lightweight MLP that directly estimates radiance on the exit surfaces. We additionally train a sampling network for effective direct illumination computation from the luminaire, and a blending network to composite the luminaire into the scene. Our formulation makes it feasible to render challenging luminaires using low sample counts in arbitrary scenes. \rev{Code is available at \url{https://github.com/pedrovfigueiredo/purelight}}.
\end{abstract}

%
%

%
%


\begin{teaserfigure}
 \centering
 \includegraphics[width=\linewidth]{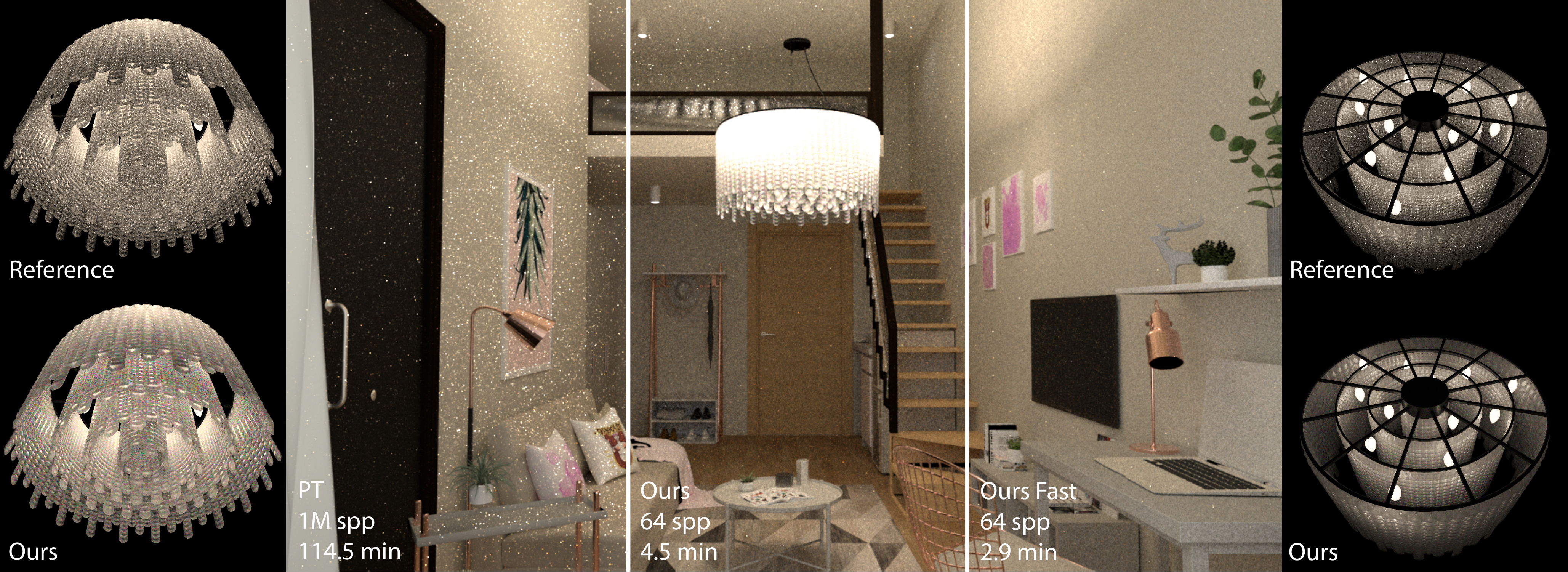}
 \vspace{-0.25in}
 \caption{We introduce PureLight, a neural formulation for estimating the appearance of complex luminaires through distribution learning from light-traced samples. As shown, our approach reconstructs luminaire appearance closely matching the reference images. Compared to path tracing with one million spp, our method produces results with lower noise using only 64 spp. We additionally introduce an alternative version of our method (Ours Fast), which is faster and produces even lower noise.}
 \label{fig:teaser}
 \end{teaserfigure}


\maketitle
\section{Introduction}
\label{sec:intro}

Physically-based Monte Carlo (MC) rendering has reached high levels of realism, but this requires high-quality scenes with physically meaningful 3D assets, materials, and lighting. Recent progress on neural representations for shapes and materials seeks to fill this need. However, light sources in most Monte Carlo rendering applications still tend to be simple: infinitely distant image-based lighting (IBLs), area lights based on simple shapes, or IES profiles of complex luminaires, which only encode far-field distributions and cannot represent complex shadows nor the luminaire appearance.

A few approaches have been proposed to precompute classical or neural representations of complex luminaires \cite{velazquez2015complex,Zhu_2021_ToG,Condor2022}, making the inclusion of highly complex luminaires in traditional rendering pipelines more practical. However, these methods still assume that standard path tracing or bidirectional path tracing can provide estimates of the luminaire appearance (when viewed directly) with reasonably low variance.  Unfortunately, this assumption does not hold for many realistic luminaire models that use detailed and physically accurate materials and emitters, such as thin glowing wires or tiny LEDs covered in multiple refractive layers of glass with low roughness. In such cases, the Monte Carlo estimation of the outgoing radiance on the surface of the luminaire can become arbitrarily difficult. Therefore, during learning of a neural luminaire model, even the problem of constructing valid training data can become a barrier to the adoption of the above methods.

We propose a neural formulation for estimating the appearance of complex luminaires with arbitrarily difficult light transport, based on \emph{distribution learning}. Our key idea is to model the probability density function (pdf) proportional to the exitant radiance of the luminaire, and recover outgoing radiance as the product of the estimated pdf and flux (for each color channel separately).  This distribution is learned purely from light-traced samples, which are straightforward to generate for arbitrary 3D luminaire geometries, regardless of complexity. Our solution takes inspiration from PureSample \cite{li2026puresample}, which constructs neural materials purely by forward path tracing of arbitrarily specular microgeometries. By constructing paths from emitters to the exit surfaces, our light tracing formulation enables learning of challenging luminaires where classical forward or bidirectional path tracing struggles to find valid light paths. To accurately represent this distribution, we decompose it into a marginal positional distribution, estimated using photon mapping, and a conditional directional distribution, learned using a large normalizing flow network.


\shorten{For faster appearance queries, we further distill the appearance into a lightweight MLP that directly estimates radiance on the luminaire exit surfaces.} Moreover, we train a blending network to composite the luminaire into any scene, similarly to \citet{Zhu_2021_ToG}.

Finally, we also train an importance sampling network based on a compact normalizing flow to support next-event estimation (direct illumination) at a given scene shading point. Additionally, we present an alternative low-resolution representation that directly encodes the appearance of the luminaire from a given shading point into a discretized $16 \times 16$ map, which can be used for both radiance evaluation and sampling, improving upon a previous solution by \citet{Zhu_2021_ToG}. This representation introduces a small bias but enables significantly lower variance, making it particularly effective in low sample regimes where slight bias is acceptable. We show that this strategy produces low-noise results at low sample counts.

In summary, the contributions of our paper are:
\begin{itemize}
    \item A distribution learning framework for precomputing a neural representation of a complex luminaire purely from \emph{light tracing} samples from the emitting components outwards, which is algorithmically straightforward even in cases where other unbiased Monte Carlo estimators of outgoing radiance have arbitrarily high variance.
    \item A solution for estimating radiance directly on the luminaire surfaces instead of a proxy bounding volume, which leads to much sharper results than \citet{Zhu_2021_ToG} without requiring UV-mapping of the luminaire surfaces.
    \shorten{\item A distillation approach that trains a lightweight network to directly estimate radiance on the luminaire surface.}
    \item An importance sampling (next-event estimation) approach that learns a sampling distribution of rays towards the luminaire, conditional on a shading point, for low-noise direct illumination estimation from the luminaire.
    \item An alternative discretized representation for evaluating and sampling the luminaire that significantly reduces the variance at the cost of slight increase in bias.
\end{itemize}


We demonstrate that our formulation makes it feasible to precompute challenging complex luminaires into an efficient neural representation and render them using low sample counts in various 3D scenes, such as the one shown in Fig.~\ref{fig:teaser}.

\section{Related Work}
\label{sec:relatedwork}

\paragraph*{Light transport methods} General MC light transport methods provide a flexible framework for rendering scenes with physically realistic lighting. Standard path tracing~\cite{10.1145/15922.15902} is unbiased and widely applicable, but it struggles to handle complex luminaires as most paths that carry substantial radiance are extremely unlikely to be sampled, resulting in high variance and slow convergence. While bidirectional path tracing~\cite{veach97} and Metropolis light transport~\cite{Veach_MLT} improve the probability of sampling rare paths, they can still fail to efficiently capture high-contribution light paths when the luminaire contains complex geometry, refractive layers, and generally specular-dominant features.

Another class of approaches, such as photon mapping~\cite{Jensen_EGSR_1995,Hachisuka_2008_TOG} and vertex connection and merging~\cite{Georgiev_2012_TOG}, trace light subpaths from the emitters, store the photons (light vertices) on scene surfaces, and treat the radiance prediction as a density estimation problem. However, these methods require a large number of photons or light vertices to avoid blurring high-frequency light features which often dominate realistic luminaires.

\paragraph*{Complex luminaires} A key limitation of all the methods above is that placing the same luminaire in different scenes requires retracing rays to simulate its internal light transport for each scene. Since most of this transport is independent of the surrounding scene, this repetition is highly inefficient for luminaires with intricate geometry or tiny emitters. Approaches in this category, which our method also belongs to, leverage this observation by precomputing or learning the luminaire’s appearance once, enabling reuse across scenes.

Early methods either measured~\cite{Ashdown1995NearFieldPM,ngai87} or simulated~\cite{Heidrich1998CannedL} the light leaving the luminaire and proposed various ways to store the resulting data. Similarly, rayset-based methods~\cite{muschaweck2011, rykowski98,mas08} record rays leaving the luminaire, but on an enclosing virtual surface (proxy). Unfortunately, these methods require representing the emission with extremely high resolution to allow high-quality direct rendering of the luminaire appearance, making them impractical in many cases.

Kniep et al.~\shortcite{Kniep_CGF_2009} use directional photon mapping, but a large number of photons are required to properly visualize the luminaire, making the radiance lookups expensive. Vel\'{a}zquez-Armend\'{a}riz et al.~\shortcite{velazquez2015complex} propose a hybrid approach, representing the luminaire illumination as a set of directional point lights, while rendering appearance using the full geometry together with precomputed radiance volumes. However, for complex luminaires, this technique requires tracing a large number of internal bounces through the luminaire, making direct rendering slow.

Condor and Jarabo~\shortcite{Condor2022} build on recent advances in 3D representations and propose encoding luminaires first as a neural radiance field (NeRF)~\cite{Mildenhall_2020}, and then as a plenoxel representation~\cite{yu2021plenoxels}. Zhu et al.~\shortcite{Zhu_2021_ToG} propose learning a lightfield on a spherical proxy together with transparency using separate MLPs, and additionally learning a pdf from arbitrary points in 3D via a discretized $16\times16$ map. A key limitation of these approaches is that training data is generated using (bidirectional) path tracing. For realistic luminaires with complex materials and emitters, sampling rare high-energy paths between a point and the emitter becomes extremely difficult. As a result, generating high-quality training data for NeRFs or MLPs becomes prohibitively expensive.

\rev{More recently, Atananov and Koylazov~\shortcite{WCL} (WCL) use light tracing samples to represent complex luminaires, accumulating photons on a bounding box proxy to approximate their illumination. Specifically, WCL discretizes the 4-dimensional domain and accumulates the photons in each bin to estimate the radiance. Using a low binning resolution, the discretization introduces significant aliasing. To combat this, WCL applies filtering, which further blurs the already low-resolution appearance. Additionally, storing this 4D discretized representation is prohibitively expensive, so WCL compresses it with a wavelet transform, further degrading quality. Although these approximations work well for estimating the illumination cast by the luminaire, they cannot faithfully represent the luminaire's direct appearance. We address this fundamental issue by proposing a strategy to estimate the appearance through distribution learning on light tracing samples. Furthermore, we learn on the luminaire’s visible surfaces rather than a proxy, disentangling directional and spatial appearance to significantly improve results.
}

\paragraph*{Neural Sampling} Several approaches have proposed the use of neural networks for importance sampling in a variety of rendering applications, including path guiding~\cite{Zheng_CGF_2019, Muller_TOG_2019, figueiredo25guidingdf, Dong_2023, Huang_Tog_2024}, light sampling~\cite{NIS_ManyLights_sig25}, and BRDF sampling~\cite{Wu_2025_SIG, Fu_2024_SIG, Xu_2023_NeuSample}. These methods employ different distribution representations and learning strategies, such as mixtures of Gaussians, histograms, flow matching, and normalizing flows, to approximate the distribution of the integrand in their respective problems. In our work, we also learn to importance sample the luminaire using a normalizing flow~\cite{Conor_2019_NIPS}, but adopt a spherical cap parameterization to minimize sampling waste.
Additionally, one of our key contributions is to formulate luminaire appearance estimation as a distribution-learning problem, which we address in part using a normalizing flow.

\section{Method}
\label{sec:method}

Our goal is to accurately render complex luminaires with arbitrarily small light emitters, enclosed in arbitrarily complex layers of (often highly specular) reflective and refractive surfaces. We design a model that learns the outgoing radiance of the luminaire using only light path samples by fitting a probability distribution proportional to the radiance (\cref{sec:appearance}). Next, we distill this large model into a smaller MLP for direct query of the radiance and we discuss how to composite the luminaire with the scene background (\cref{sec:distill}). Then, we propose a discretization-free strategy to efficiently sample direct illumination from the luminaire (\cref{sec:sampling}), and present an alternative importance sampling strategy that exchanges bias for faster runtime (\cref{sec:fast_sampling}).

\subsection{Preliminaries}

We can express directions as 3D unit vectors $\omega$ on the unit sphere $\Omega$. We define a luminaire as a surface manifold $\manifold$.  Points $\xx \in \manifold$ have corresponding normals $\n(\xx)$. The outgoing radiance on any point $\xx$ in direction $\diro$ satisfies the rendering equation \cite{10.1145/15922.15902}:
\begin{equation}
\label{eq:rendering_eq}
    L_o(\xx, \diro) = L_e (\xx, \diro) + \int_\Omega \bsdfsym(\xx, \diro, \diri) \ L_i(\xx, \diri) \ \text{d}\sigma_\bot(\diri),
\end{equation}
where $\bsdfsym$ is the bidirectional scattering distribution function (BSDF), $L_i$ is incoming radiance and $L_e$ is emitted radiance. Generally, $L_e$ will be zero except on the actual emitter elements of the luminaire (filaments, LEDs, etc.). Finding $L_o(\xx, \diro)$ using classical Monte Carlo estimators can become arbitrarily difficult as there is no limit to how small the emitter can be, and how many (often highly specular) surfaces enclose it. Our goal is to learn the appearance of the luminaire, which is the outgoing radiance $L_o(\xx, \diro)$ of surface points $\xx \in \surf$, where $\surf \subseteq \manifold$ is the subset of luminaire surfaces that are visible to the rest of the scene. For exposition simplicity, we do not consider volumetric effects (absorption and scattering) within dielectric parts of the luminaire, though they are easy to add in practice and our implementation supports them.

We will make use of the projected solid angle measure $\sigma_\bot(\diro)$, which measures a set of directions as the area of their projection on the unit disk (projected hemisphere), as opposed to the solid angle measure $\sigma(\omega)$, which measures a set of directions as their area on the unit hemisphere. We have $\dd \sigma_\bot(\diro) = |\cos \theta| \ \dd \sigma(\diro)$, and for pdfs we have $p_{\mbox{sa}}(\diro) = |\cos \theta| \ p_{\mbox{psa}}(\diro)$, where $\theta$ is the angle between $\diro$ and the surface normal $\n(\xx)$.
All pdfs we will consider are taken with respect to the projected solid angle measure unless noted otherwise.

For simplicity, assume a grayscale (single wavelength) luminaire; we will get rid of this assumption later.

\subsection{Learning Appearance with Light Tracing}
\label{sec:appearance}

Constructing an efficient estimator for $\Lo(\xx, \diro)$ from light-tracing samples is challenging. \rev{While binning the samples over position and direction is a potential solution, obtaining a high-resolution estimate is impractical, whereas using a low-resolution representation, as in WCL~\cite{WCL}, is inadequate for faithfully reproducing direct luminaire appearance.} Our key idea is to address this problem through distribution learning and obtaining an approximation of the pdf $\pdf(\xx, \diro)$ which (for a given wavelength or color channel of interest) is proportional to the outgoing radiance $\Lo(\xx, \diro)$ as follows:

\begin{equation}
    \Lo(\xx, \diro) = \flux \cdot \pdf(\xx, \diro),
    \label{eq:radiance_recovery}
\end{equation}

\noindent where $\flux$ is the proportionality constant and is simply the total flux of the luminaire, whose approximation, $\appflux$, can be easily computed. Note the convenience of the pdf being in the projected solid angle measure: if the luminaire is simply a Lambertian emitter, its radiance as well as the pdf are constants, while in the solid angle measure, the pdf would be a directional cosine term.

To obtain a neural approximation $\pdf_\theta(\xx, \diro)$ of the target distribution, we minimize the Kullback-Leibler (KL) divergence between $\pdf(\xx, \diro)$ and the learned $\pdf_\theta(\xx, \diro )$ as follows:

\begin{equation}
    D\big(\pdf \ \Vert \ \pdf_\theta\big) = \int_\surf \int_\hemi \pdf(\xx, \diro) \log \frac{\pdf(\xx, \diro)}{\pdf_\theta(\xx, \diro)} \ \text{d}\sigma_\bot(\diro) \ \text{d}\xx.
\end{equation}

\noindent Minimizing $D$ necessitates computing its gradient with respect to the network parameters $\theta$, as follows:

\begin{equation}
    \nabla_\theta  D\big(\pdf \ \Vert \ \pdf_\theta\big) = - \int_\surf \int_\hemi \pdf(\xx, \diro) \nabla_\theta\log \pdf_\theta(\xx, \diro) \ \text{d}\sigma_\bot(\diro) \ \text{d}\xx.
\end{equation}

Using Eq.~\ref{eq:radiance_recovery}, we can replace $\pdf$ with $\Lo$ as the flux is a constant and only adjusts the scale of the gradient. Writing the path integral formulation of $\Lo$, we have:

\begin{equation}
\nabla_\theta D\big(\pdf \ \Vert \ \pdf_\theta\big)
= - \int_\surf \int_\hemi
\ \underbrace{\int_{\Omega(\xx, \diro)}\!\! f(\xb)\,\text{d}\xb}_{\Lo(\xx, \diro)}
\ \nabla_\theta \log \pdf_\theta(\xx, \diro)
\ \text{d}\sigma_\bot(\diro) \ \text{d}\xx,
\end{equation}

\noindent where $f(\xb)$ is measurement contribution evaluated on path $\xb$ from one of the emitters to an exit surface. Furthermore, the integral is over the space of all possible paths $\Omega(\xx, \diro)$ that terminate at $(\xx, \diro)$. We can further collapse the three integrals into one integral over all the paths (terminating at any $\xx, \diro$) as follows:

\begin{align}
\label{eq:finalgrad}
    \nabla_\theta  D\big(\pdf \ \Vert \ \pdf_\theta\big) &= - \int_{\Omega} f(\xb) \nabla_\theta\log \pdf_\theta(\xx, \diro) \ \text{d}\xb \\ \nonumber
    &= \mathbb{E} \big[ - \frac{f(\xb)}{q(\xb)} \nabla_\theta\log \pdf_\theta(\xx, \diro) \big],
\end{align}

\noindent where the expectation is over paths $\xb$ drawn from pdf $q$.

This is an importance-weighted negative log-likelihood objective, which can be efficiently optimized using light-traced paths. Specifically, we begin by sampling the union of all emitter surfaces, then continue tracing the path through interactions with luminaire surfaces until it exits to infinity. The path contribution (emitted radiance, BSDF, and geometry terms), with the path sampling density, is used to construct the importance weight (path throughput) $f(\xb)/q(\xb)$. We additionally record the final surface interaction $(\xx,\diro)$, consisting of the exit position and direction, which is used to evaluate the log-likelihood $\log \pdf_\theta(\xx,\diro)$. Next, we discuss our approach for modeling $\pdf_\theta$.

\subsubsection{Modeling through Distribution Factorization}
\label{sec:appearance_fact}

Instead of learning the 4D distribution $p(\xx, \diro)$ directly, using a 4D normalizing flow, we factorize the joint distribution into a marginal spatial distribution $\pdfPos$ over the exit surfaces $\surf$ and a conditional directional distribution $\pdfDir$ as follows:
\begin{equation}
    \pdf_\theta(\xx, \diro) = \pdfPos(\xx) \cdot \pdf_\theta(\diro \mid \xx).
    \label{eq:factorization}
\end{equation}

This factorization has a major advantage over the direct 4D approach. The point $\xx \in \surf$ is really a 3D point on a 2D manifold; to apply normalizing flows to sampling $\xx$ directly, we would need a high quality UV parameterization of $\surf$, which is not always easy to obtain. 
\shorten{With this factorization, we estimate $\pdfPos$ using simple density estimation and learn only $\pdf_\theta(\diro \mid \xx)$ with a normalizing flow; $\xx$ only conditions the flow, requiring no 2D parameterization.}

We note that an alternative solution is to learn the distribution on a proxy geometry, e.g., a bounding sphere. However, compared to learning on the actual geometry, this strategy leads to inferior results as the spatial and directional appearance become entangled and more difficult to learn (see Sec.~\ref{sec:ablations}).

\begin{figure}[t]
\centering
\includegraphics[width=\linewidth]{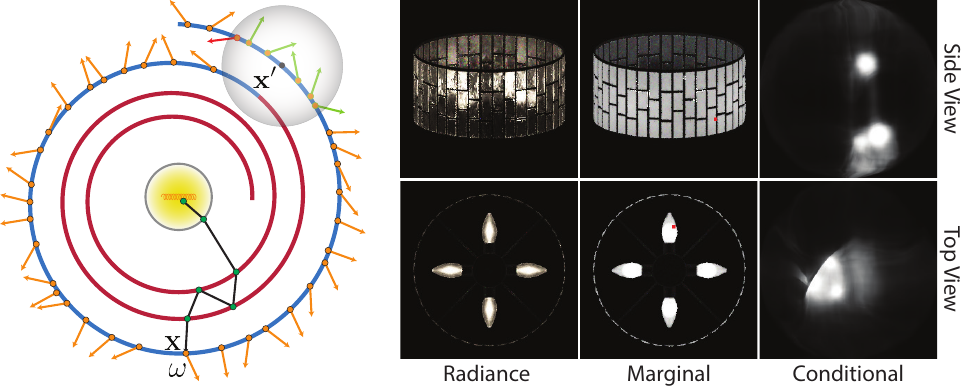}
\caption{On the left, we illustrate a luminaire consisting of an emissive filament enclosed within a circular glass bulb, surrounded by a refractive spiral glass enclosure. We perform light tracing starting from the emitter, with paths bouncing through the luminaire until they exit. At the last surface interaction, we record the sample position and exit direction $(\xx, \diro)$, which are used to learn a distribution $\pdf(\xx, \diro) = \pdfPos(\xx) \cdot \pdf(\diro \mid \xx)$ proportional to $\Lo(\xx, \diro)$. The black path illustrates a representative light trajectory from the filament to the recorded sample $(\xx, \diro)$. Moreover, we highlight in blue the subset of luminaire surfaces that are visible to the outside; samples are recorded only on these surfaces. We learn $\pdf(\diro \mid \xx)$ from the stored samples using normalizing flows. To estimate the spatial density at a surface point $\xx'$, $\pdfPos(\xx')$, we perform density estimation over photons within a fixed radius of $\xx'$, considering only those whose directions lie within the hemisphere defined by the surface normal. Photons indicated with green borders and arrows are included, while the red one is excluded as it lies outside the hemisphere. On the right, we show our estimated marginal and conditional distributions, along with the corresponding radiance for two views of the \textsc{Medieval} luminaire. The condition $\xx$ for the conditional distribution is visualized with a red dot in the marginal images.}
\label{fig:light_tracing}
\end{figure}

\paragraph{Estimating the positional distribution} We approximate $\pdfPos(\xx)$ using a simple density estimation approach, used in classical photon mapping \cite{Jensen_EGSR_1995}. Specifically, to estimate the pdf at surface point $\xx$, we first identify the particles within a neighborhood of radius $r$, whose directions are in the hemisphere defined by the local shading normal $\n(\xx)$. An illustration of this process can be seen in Fig.~\ref{fig:light_tracing}. We then compute the positional pdf at $\xx$ as follows:

\begin{equation}
    \pdfPos(\xx) = \frac{1}{\appflux}\sum_{i \in \mathcal{N}(\xx, r)} \frac{f(\xb_i)}{q(\xb_i)}\frac{1}{N \pi r^2},
    \label{eq:positional}
\end{equation}

\noindent \shorten{where $\mathcal{N}(\xx, r)$ denotes the set of particles neighboring $\xx$, as described above, and $N$ is the total number of emitted particles.}

\paragraph{Learning the directional distribution.} The remaining core problem is how to learn $\pdf(\diro \mid \xx)$ using a neural approximation $\pdf_\theta(\diro \mid \xx)$. We use a normalizing flow architecture with piecewise-quadratic coupling layers \cite{Muller_TOG_2019} and train it using the objective in Eq.~\ref{eq:finalgrad}. Note that here, we only need to compute the log likelihood of the directional distribution $\pdf_\theta(\diro \mid \xx)$ as the derivative of $\pdfPos$ with respect to $\theta$ is zero.

We represent the directions $\diro$ in the local shading frame using a unit disk parameterization. To condition the model, we provide the 3D position $\xx$ normalized by the bounding box and encoded with a learnable hash grid~\cite{Muller_TOG_2022}. We additionally provide the surface normals, encoded as 2D spherical coordinates using one-blob encoding~\cite{Muller_TOG_2019}. 
\shorten{We use $8$ coupling layers, each with five 256-neuron hidden layers and ReLU activations.}

\paragraph{Supporting RGB color.} We assumed a grayscale luminaire until now. To support RGB colors (or multiple wavelengths), we simply fit to the color channels independently. Our positional distribution estimation method in Eq.~\ref{eq:positional} naturally produces three $\pdfPos$ estimates per channel. In the optimization of the directional distribution, we add an extra conditioning to the model representing a color channel, which we encode using one-hot encoding. To compute the RGB radiance, we compute the per-channel product with the flux (Eq.~\ref{eq:radiance_recovery}). A visualization of each component of our learned representation can be seen in  Fig.~\ref{fig:light_tracing}.

\begin{figure}[t]
\centering
\includegraphics[width=0.9\linewidth]{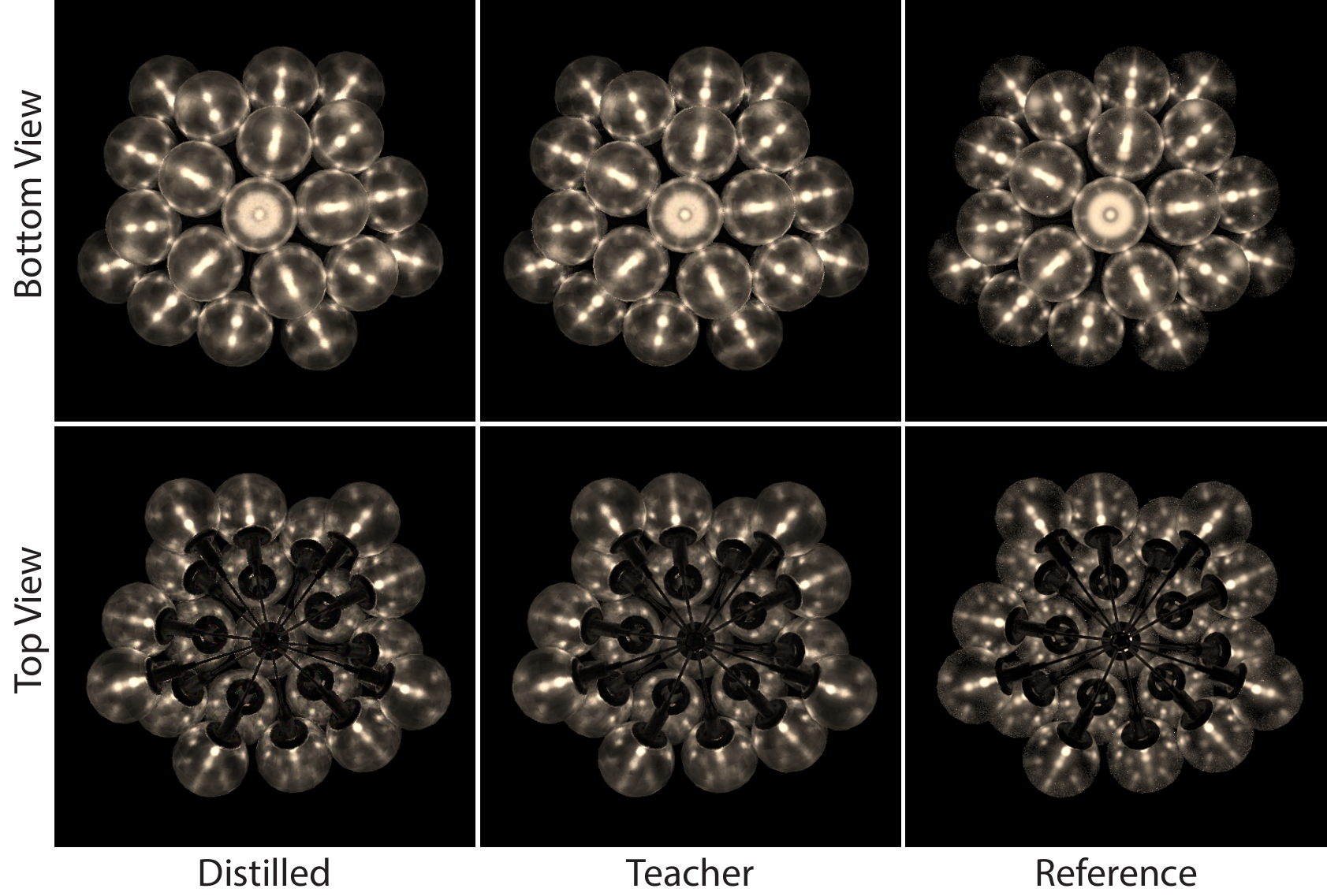}
\vspace{-0.1in}
\caption{We show a comparison of our appearance estimation using a large normalizing flow (Teacher), the distilled lightweight network (Distilled), and the reference. As seen, the distilled network produces results of similar quality to the teacher, and both are reasonably close to the reference.}
\label{fig:distill_comparison}
\end{figure}

\subsection{Appearance Distillation and Transparency}
\label{sec:distill}

\paragraph{Distillation.} To enable faster rendering, we distill the estimated appearance, the product of photon mapping and large normalizing flow, into a smaller network $\netStudent(\xx, \diro)$ that directly regresses radiance on the surface of the luminaire. Unlike \citet{Zhu_2021_ToG} who learn the appearance on a bounding proxy, we condition our distilled network on the luminaire mesh surface points $\xx \in \surf$, enabling higher fidelity appearance, as seen in Sec.~\ref{sec:ablations}. Specifically, we encode surface points $\xx$ using a learnable hash grid~\cite{Muller_TOG_2022}, accompanied by the outgoing directions $\diro$ and surface normals, both encoded with spherical harmonics of degree five. To ensure high-fidelity distillation, we use an MLP with 5 hidden layers of 512 neurons each and ReLU activations.


To optimize this student network $\netStudent(\xx, \diro)$, we collect training samples by intersecting rays with the luminaire mesh and querying the teacher appearance model. We construct rays by uniformly sampling points on the spherical bounds of the luminaire and choosing uniform directions on the hemisphere toward the luminaire center. For our objective, we follow~\citet{Zhu_2021_ToG} in computing the $\mathcal{L}_2$ loss in $\log(1+x)$ domain to better handle HDR radiance targets. 
\shorten{Fig.~\ref{fig:distill_comparison} compares the teacher and distilled student appearances.}

\paragraph{Transparency Network.} To be able to composite the luminaire onto any scene, we also train a transparency (blending) network, following~\citet{Zhu_2021_ToG}. Specifically, we trace rays toward the luminaire tracking the transmission throughput of paths ignoring changes of direction due to refraction. For this simpler task, we reduce the complexity of our appearance distillation network encodings from spherical harmonics degree five to three, while also reducing the number of neurons to 128. We use sigmoid as the output activation, and compute the $\mathcal{L}_2$ during optimization.


\subsection{Importance Sampling for Direct Illumination}
\label{sec:sampling}

\begin{figure}[t]
\centering
\includegraphics[width=\linewidth]{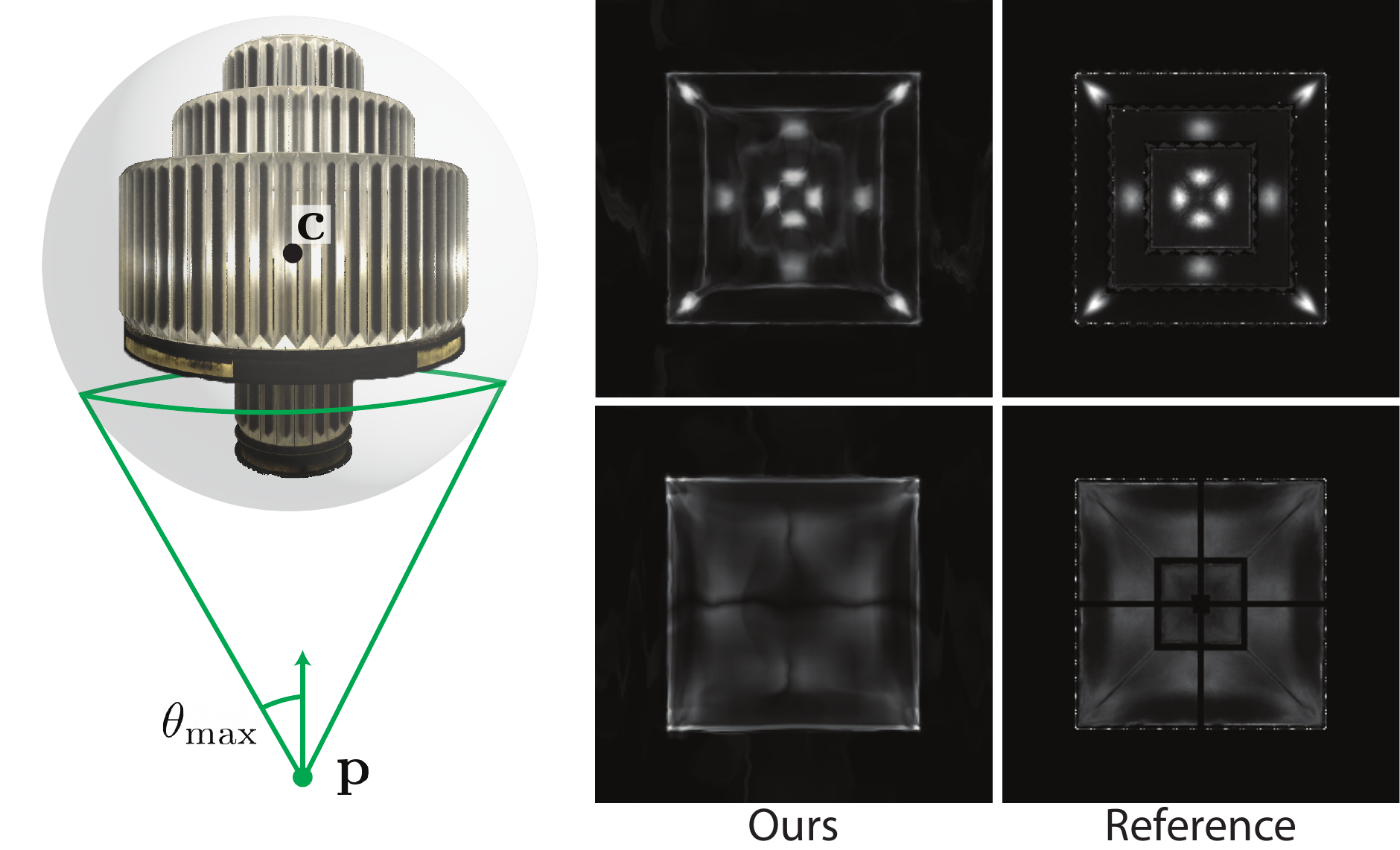}
\vspace{-0.25in}
\caption{
On the left, we show the spherical cap parameterization used for importance sampling. On the right, we show a comparison between our learned and reference distributions from shading points $\pp$ on the top and bottom of the luminaire.}
\label{fig:IS_param}
\end{figure}

The above distribution learning approach and distillation succeed in learning the appearance of the luminaire for fast direct viewing. A key remaining issue is to support importance sampling of the luminaire for next-event estimation (direct illumination). Ideally, we would like to sample a point $\xx$ on the luminaire with a probability proportional to $L_o(\xx, -\dd)$ where $\dd$ is the direction from $\pp$ towards $\xx$. Similar to the appearance model, we learn $\pdf_\phi(\dd \mid \pp)$ using a normalizing flow architecture conditioned on the shading point $\pp$. To provide tight coverage of the luminaire, we first compute the bounding sphere and parameterize the direction $\dd$ in uniform spherical cap domain, as shown in Fig.~\ref{fig:IS_param}.

\paragraph{Network}
Similar to the appearance model, we use a normalizing flow architecture with piece-wise quadratic coupling layers. 
\shorten{However, we reduce the network to $4$ coupling layers, each comprising a 5-layer MLP with 128 neurons per layer and ReLU activations.}
We provide the shading point $\pp$ and the spherical-cap half-angle $\theta_{\max}$ as the input conditions to the network. Instead of directly feeding $\pp$ to the model, we first normalize it as $(\pp - \textbf{c})/r_{\max}$ to ensure the values are in range $[-1, 1]$. Here, $\textbf{c}$ denotes the bounding sphere center and $r_{max}$ (15 in our implementation) is the maximum distance of the training samples $\pp$ to the sphere center.  We encode this normalized 3D vector with a learnable dense grid encoding~\cite{Dong_2023}, and encode $\theta_{\max}$ with one-blob encoding~\cite{Muller_TOG_2019}.

\paragraph{Optimization} 
We optimize this network by minimizing the KL divergence between the target and learned pdf. Since the target pdf is not available, we use radiance estimates as an unnormalized approximation of the target pdf~\cite{Muller_TOG_2019, NIS_ManyLights_sig25}. To generate the training data, we uniformly sample shading points in the volume between the luminaire’s bounding sphere and a sphere of radius $r_{\max}$. At each shading point, we then uniformly sample the direction $\dd$ within the spherical cap. We query the radiance corresponding to each shading point $\pp$ and direction $\dd$ from our distilled appearance network and convert RGB to luminance. These triplets form the training data for the normalizing flow network. 

\subsection{Fast Importance Sampling}
\label{sec:fast_sampling}

Although the importance sampling strategy in Sec.~\ref{sec:sampling} yields a discretization-free solution to sample and evaluate the PDF of a luminaire radiance field, it still requires querying the distilled appearance network to retrieve radiance samples used during next-event estimation (NEE). This results in additional computational overhead, requiring two sequential network inferences at each NEE evaluation. It can also result in a mismatch between the estimated pdf and the evaluated radiance, leading to some variance. 

An alternative solution is to approximate a discrete 2D radiance map proportional to the outgoing radiance from the luminaire $L_o(\xx, -\dd)$ conditioned on the shading point $\pp$, as proposed by~\citet{Zhu_2021_ToG}. In this way, the network's output can be used for both sampling a direction and evaluating the radiance. The result is a faster NEE computation with low variance, but with a slight bias due to the low-resolution radiance approximation. \rev{We present the results of this strategy as our \textit{fast} approach. Note that this strategy is strictly for modeling illumination from the luminaire. For the direct appearance from camera rays, we still query the distilled appearance and transparency networks.}

\paragraph{Network and Optimization}
We use a $512\times 5$ MLP with ReLU activations and a similar input encoding to the normalizing flow architecture in Sec.~\ref{sec:sampling} with an extra input to accommodate the channel conditioning encoded with one-hot encoding. Following \citet{Zhu_2021_ToG}, we estimate $16 \times 16$ radiance maps for all our results, but instead of using perspective camera, we use the spherical cap parametrization for a tighter coverage of the luminaire. We optimize the network with a $\mathcal{L}_2$ loss, so the model focuses more on the higher energy pixels. However, we still ensure that the network produces the output in the $\log(1+x)$ domain to properly handle HDR values.

\section{Results}
\label{sec:results}

In this section, we present comparisons of our method with multiple luminaires placed in various scenes to demonstrate its effectiveness. Specifically, we benchmark our approach against path tracing (PT) in Sec.~\ref{sec:ptcomp}, bidirectional path tracing (BDPT) in Sec.~\ref{sec:bdptcomp}, and NCL~\cite{Zhu_2021_ToG} (Sec.~\ref{sec:nclcomp}). We use the \rev{same GPU-based} Mitsuba 3 renderer~\cite{Mitsuba3} for our data generation and rendering, and for rendering PT and NCL. As BDPT is not featured in this version of Mitsuba, we use Mitsuba 0.5~\cite{Mitsuba3} to generate its renderings. We model our networks with tiny-cuda-nn~\cite{tiny-cuda-nn} and optimize them using PyTorch~\cite{pytorch}. For all methods, we terminate rendering paths probabilistically with Russian Roulette starting after the fifth bounce. All experiments and comparisons are performed on a workstation with an AMD Ryzen 5 5800X CPU and an NVIDIA RTX 4090 GPU.

\subsection{Comparison with Path Tracing}
\label{sec:ptcomp}

\begin{figure}[t]
\centering
\includegraphics[width=\linewidth]{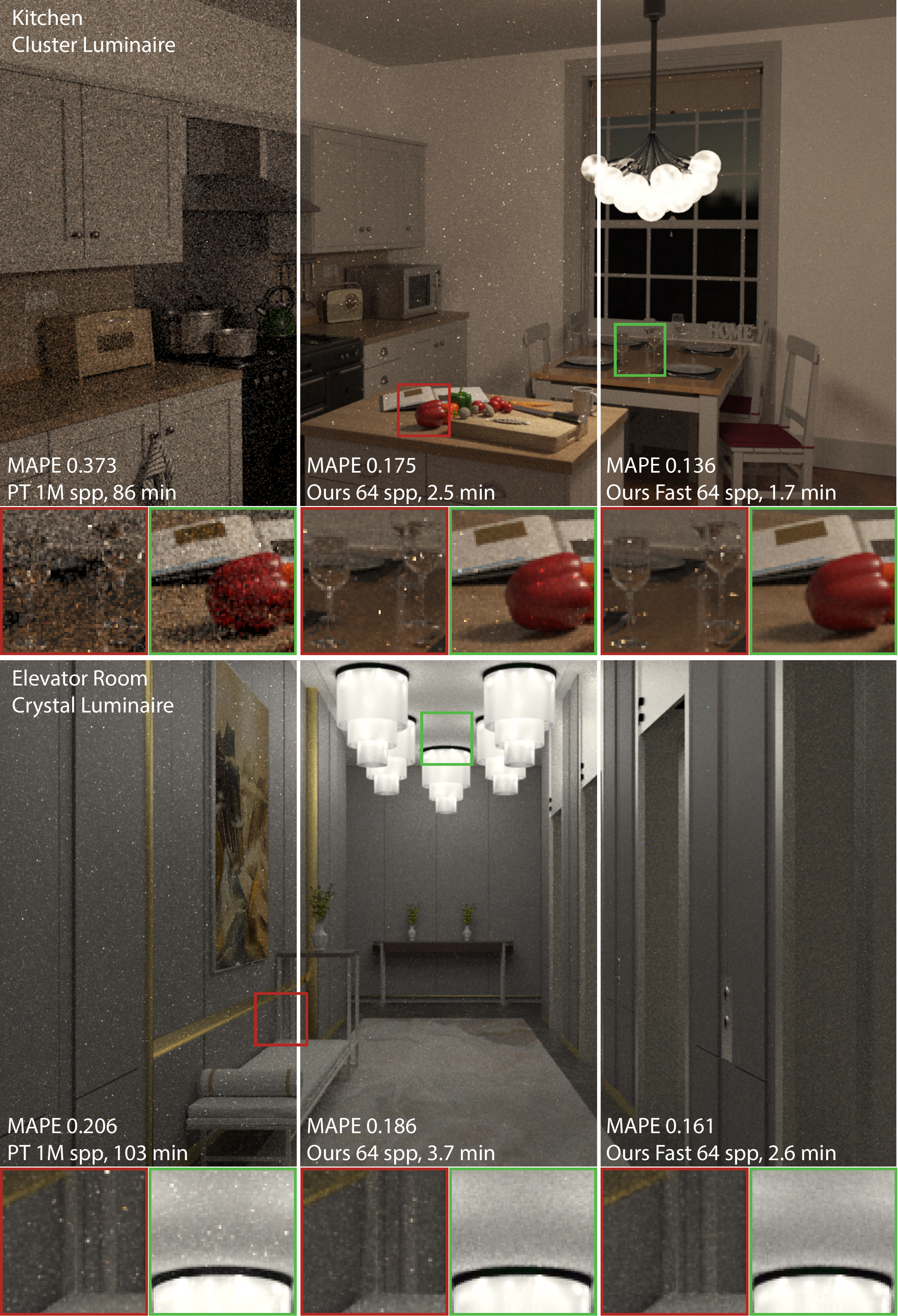}
\vspace{-0.25in}
\caption{Comparison of our method against path tracing (PT) with NEE. We report numerical error in terms of mean absolute percentage error (MAPE).}
\label{fig:pt}
\end{figure}

We compare both variants of our approach against path tracing in Fig.~\ref{fig:pt}. In the \textsc{Kitchen} scene, we render a multi-lobe \textsc{Cluster} luminaire whose light sources are small filaments inside layered glass bulbs. In the \textsc{Elevator Room}, we place multiple instances of the highly refractive \textsc{Crystal} luminaire on the ceiling of an elevator hall. The instances are all derived from the same learned representation under various transformations and therefore do not increase memory requirements. With only 64 samples per pixel (spp), our method renders the scenes with lower noise than path tracing with next-event estimation (PT) at one million spp. Our fast variant is more than 40\% faster than the normalizing flow variant and produces renderings with visibly lower noise and fewer fireflies, while introducing little perceived bias.

We note that our prototype implementation is constrained by the limited support for neural network inference within rendering loops in Mitsuba. Specifically, evaluating our neural networks requires disabling symbolic execution during rendering, contributing to a significantly higher runtime. 
\shorten{For instance, disabling symbolic execution increases the path tracing time for the one million spp \textsc{Kitchen} rendering from 86 minutes to 88 hours, a $\sim 60\times$ increase.}

\subsection{Comparison with Bidirectional Path Tracing}
\label{sec:bdptcomp}

\begin{figure}[t]
\centering
\includegraphics[width=\linewidth]{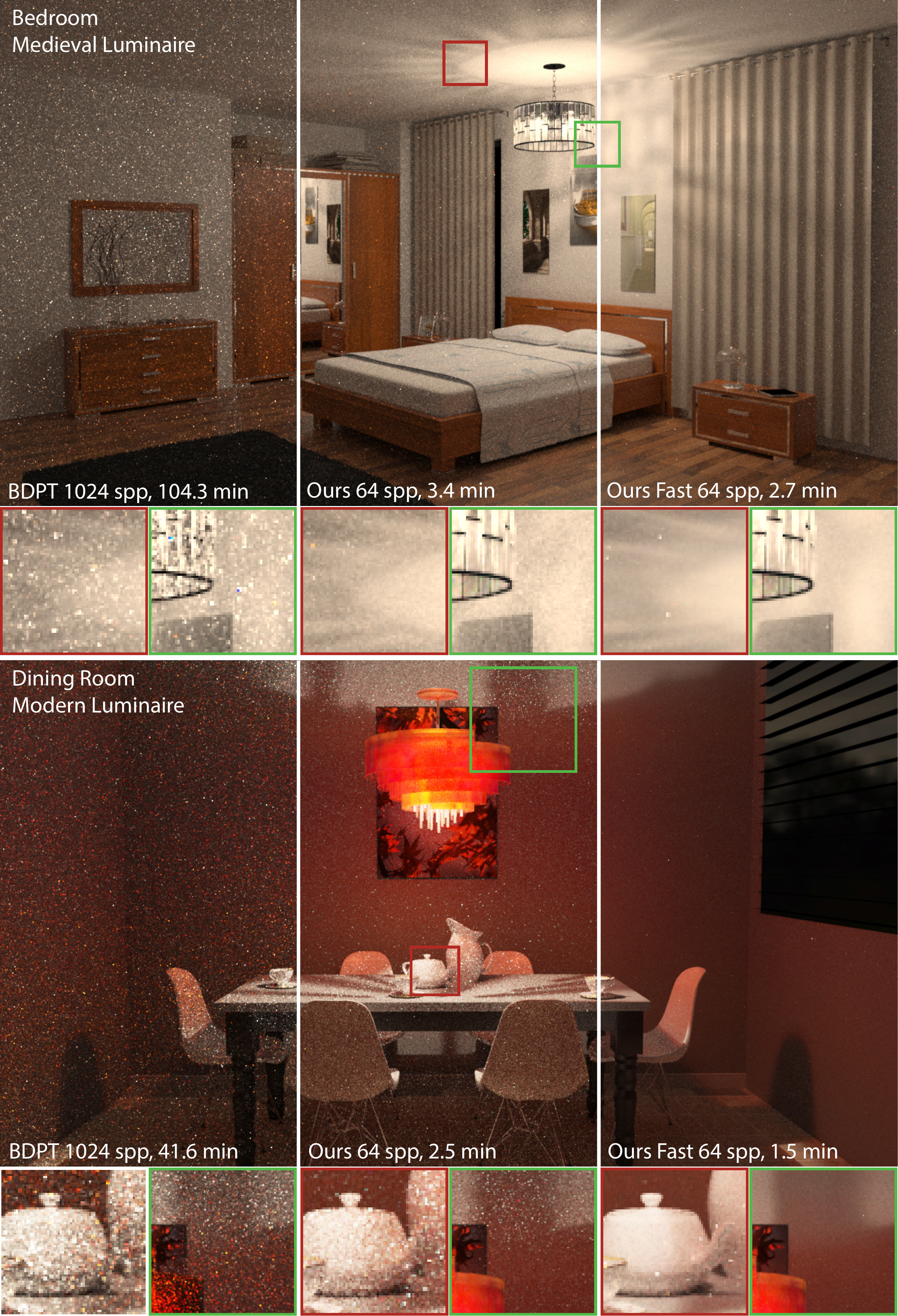}
\vspace{-0.25in}
\caption{Comparison of our method against BDPT on scenes containing challenging luminaires.}
\label{fig:bdpt}
\end{figure}

We compare our method with bidirectional path tracing (BDPT) in Fig.~\ref{fig:bdpt}. In the \textsc{Bedroom} scene, the hollowed \textsc{Medieval} luminaire projects sharp patterns on the walls and ceiling due to its textured glass surface. In the \textsc{Dining Room}, a layered \textsc{Modern} luminaire illuminates the scene with a single bright bulb enclosed by crystals. In both luminaires, the light emitters are filaments enclosed by layers of alternating glass and diffuse surfaces, resulting in many specular-diffuse-specular (SDS) light paths which are notoriously difficult to handle. As seen, our method produces results with significantly less noise at a fraction of the samples used by BDPT. Our fast variant produces renderings with considerably less noise in a shorter time, but shows slight bias in the soft shadows produced by the inner crystals of the \textsc{Dining Room} scene (see green inset). 

We note that we are not able to report numerical errors here, as these luminaires exhibit extremely complex light transport, making the computation of ground-truth images using path tracing impractical. Comparison against a reference image obtained using BDPT is also not meaningful due to differences in shading caused by the use of a different version of Mitsuba for BDPT.

\subsection{Comparison with NCL}
\label{sec:nclcomp}

\begin{figure}[t]
\centering
\includegraphics[width=\linewidth]{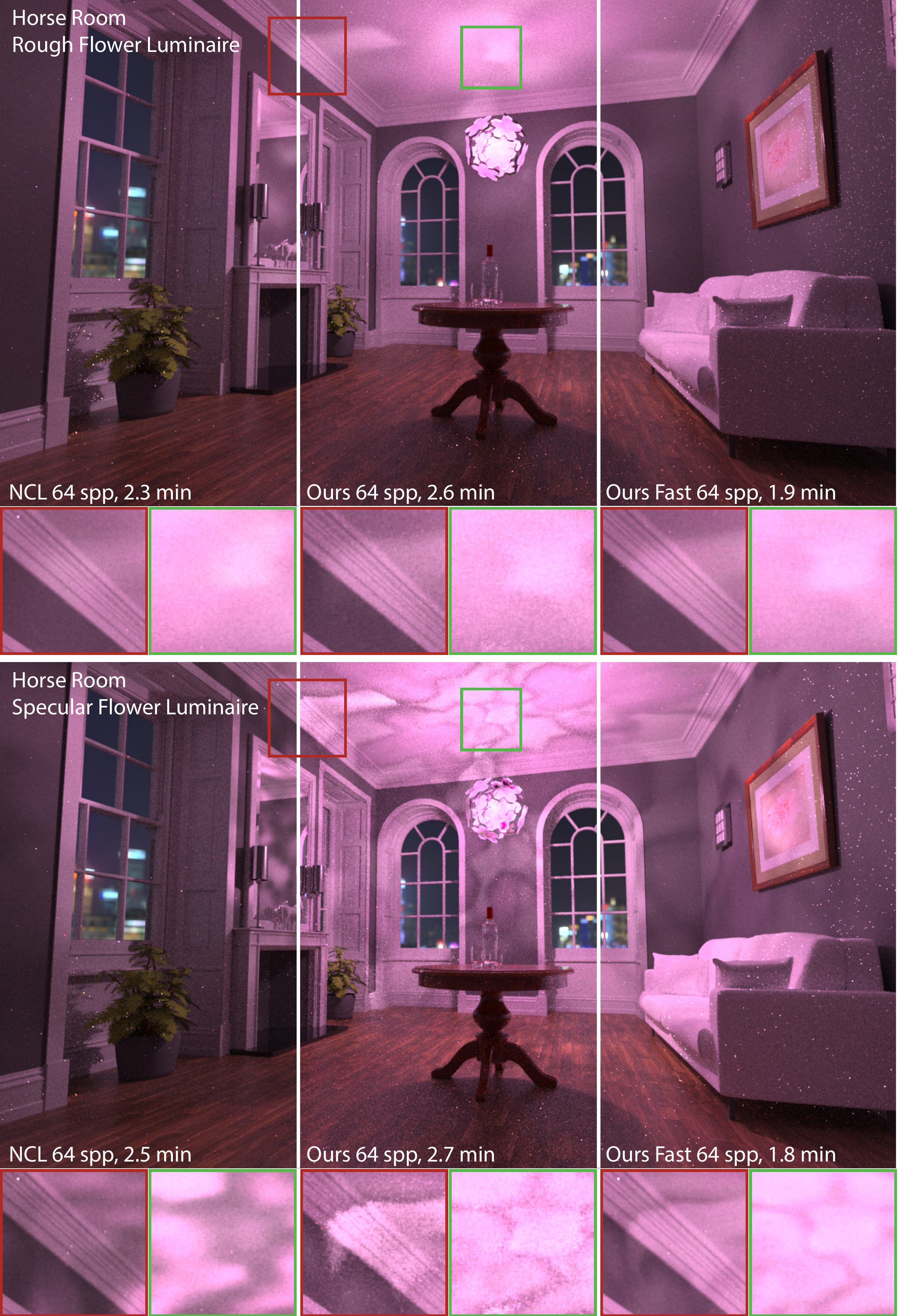}
\vspace{-0.25in}
\caption{Comparison of our method against NCL~\cite{Zhu_2021_ToG} on two versions of the \textsc{Flower} luminaire with rough (top) and highly specular (bottom) glass layers. While NCL produces reasonable results for the simpler rough flower, it exhibits significant bias for the specular flower, as generating the training images becomes challenging. Our method, however, is able to handle both cases.}
\vspace{-0.0in}
\label{fig:nclcomp-flower}
\end{figure}

\begin{figure}
\centering
\includegraphics[width=\linewidth]{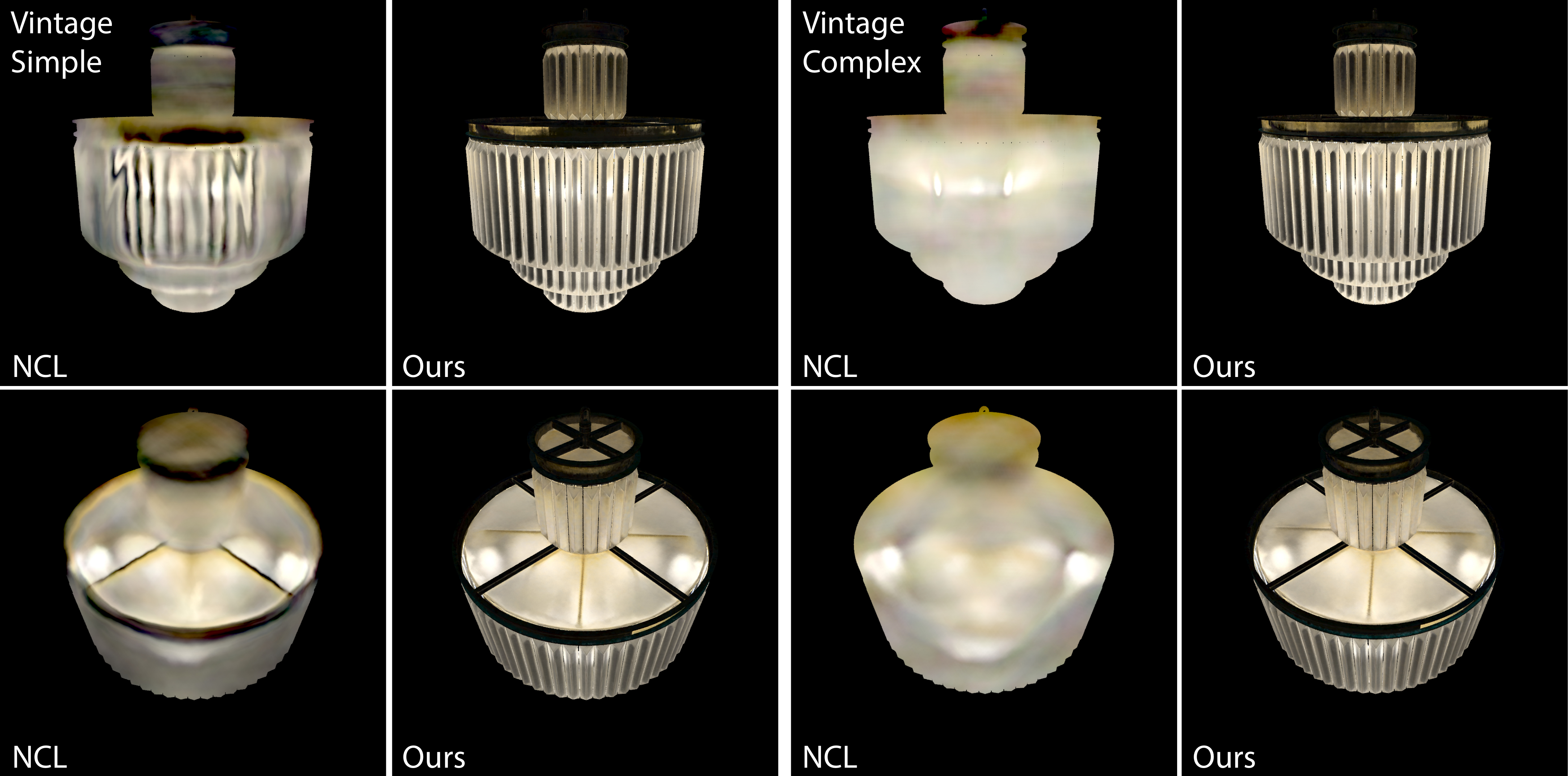}
\vspace{-0.25in}
\caption{Comparison of our method against NCL~\cite{Zhu_2021_ToG} for direct rendering of two versions of the \textsc{Vintage} luminaire.}
\label{fig:nclcomp-glasschandelier}
\end{figure}

We compare our approach with NCL~\cite{Zhu_2021_ToG} in Figs.~\ref{fig:nclcomp-flower} and \ref{fig:nclcomp-glasschandelier}. Each figure includes comparisons on two luminaires: a simpler luminaire similar to those shown in the original paper, and a luminaire with significantly more complex light transport that is challenging for NCL due to its reliance on BDPT for constructing training data. As the complexity of light transport increases, the BDPT-generated training targets used by NCL exhibit higher variance, which can degrade the quality of the learned representation.

The \textsc{Rough Flower} luminaire (Fig.~\ref{fig:nclcomp-flower} - top) features multiple layers of flower petals made of colored glass with a small emissive filament in its center. \textsc{Specular Flower} shares the same geometry and light while having flower petals with lower roughness. As seen, decreasing the flower petal roughness transforms the projected illumination across the scene from smooth and blurry patterns to sharper contours of each petal. NCL correctly predicts the illumination patterns for the \textsc{Rough Flower}, but introduce\rev{s} severe bias for the specular version. In contrast, our method based on light tracing is able to handle both cases effectively. Although considerably faster, our fast variant also suffers from a slight increase in bias for the \textsc{Specular Flower}. 
\shorten{This is expected when representing peaky radiance profiles at a fixed low resolution.}

In Fig.~\ref{fig:nclcomp-glasschandelier}, we compare our approach against NCL on direct rendering of two luminaires. The \textsc{Vintage Simple} luminaire (top) features a closed rough glass and gold surface inside of which many bulbs are distributed uniformly. \textsc{Vintage Complex} shares geometry and materials, but changes the complexity of the bulbs from larger meshes to more realistic filaments inside glass enclosures. Although blurrier than ours, NCL produces reasonable results for the simpler luminaire. However, for the complex one, NCL is not able to capture the sharp details due to increased variance in its training data. Our method produces sharper estimates regardless of the complexity of light bulbs inside of the luminaire.

\rev{
\subsection{Pre-computation Time and Memory}
\label{sec:timemem}
Table~\ref{tab:precomputation_time} summarizes the pre-computation time for each stage of our method. Distillation includes 5--8 hours of data generation because, for simplicity, our implementation evaluates densities on the CPU, but this process could be accelerated using GPU implementations in the future. For the other stages, samples are generated during optimization, whose runtime is dominated by network training. During rendering, the learned representations occupy 343\,MB for our method and 246\,MB for our \textit{fast} variant. These totals comprise 90\,MB for distilled appearance, 73\,MB for transparency, and 180\,MB and 93\,MB, respectively, for importance sampling.
}

\begin{table}[t]
\setlength\tabcolsep{3.5pt}
\centering
\caption{\rev{Pre-computation times, reported in hours. Stages separated by ``+'' run in parallel.}}
\vspace{-0.1in}
\label{tab:precomputation_time}
\footnotesize
\begin{tabular}{lccc}
\toprule
Stage & Data Gen. & Optimization & Total \\
\midrule
Teacher (Photon Accum. + Directional NF)
    & -- & 10--16 & 10--16 \\
Distillation
    & 5--8 & 1 & 6--9 \\
Importance Sampling + Transparency
    & -- & 3--6 & 3--6 \\
\midrule
\textbf{Full Algorithm}
    & \textbf{5--8} & \textbf{14--17} & \textbf{19--23} \\
\bottomrule
\end{tabular}
\end{table}

\subsection{Ablations}
\label{sec:ablations}

\begin{figure}
\centering
\includegraphics[width=\linewidth]{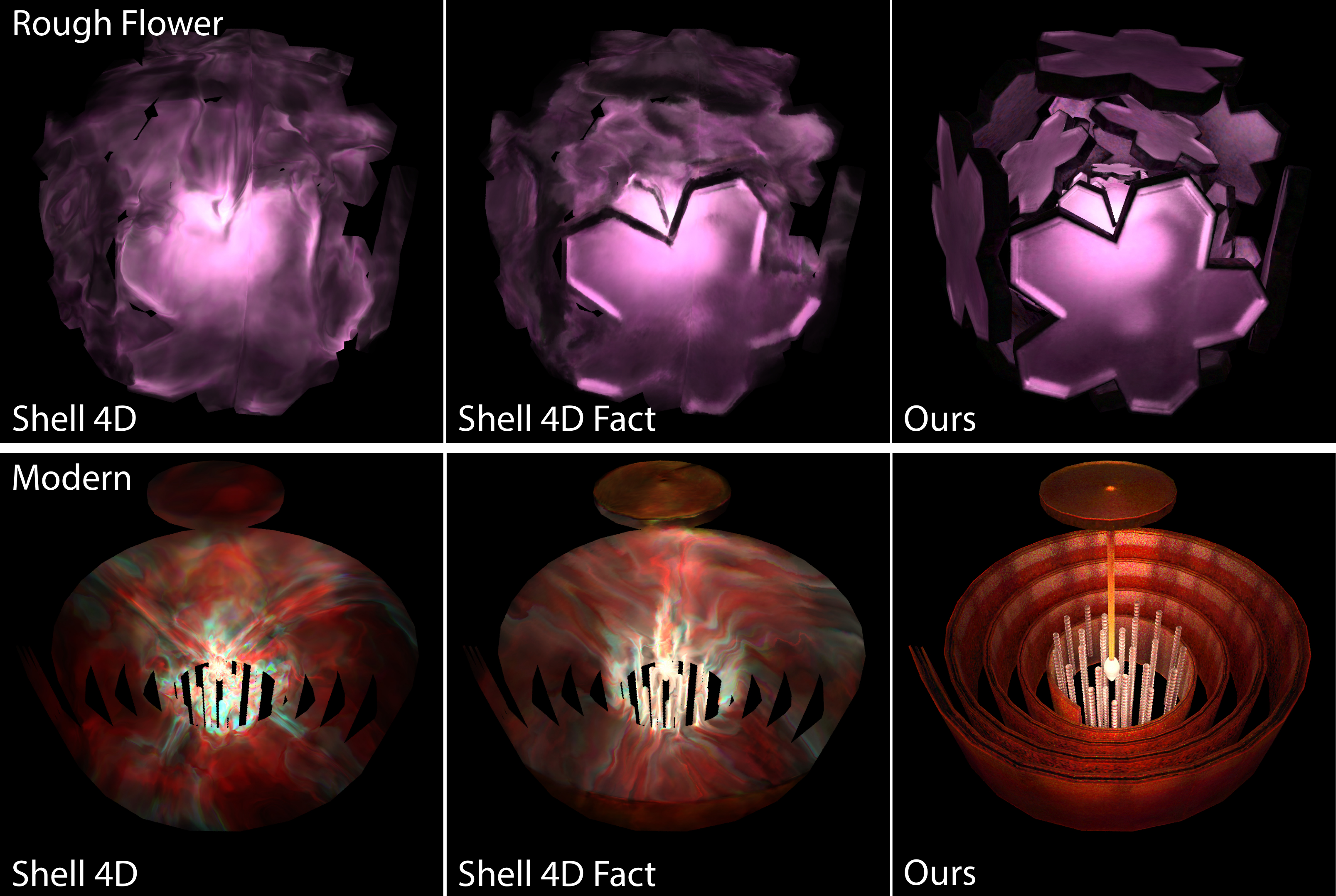}
\vspace{-0.25in}
\caption{Comparison of our approach, which learns appearance on the exit surfaces, against two variants based on a spherical proxy geometry: directly learning the 4D distribution (Shell 4D) and learning separate marginal and conditional distributions on the shell (Shell 4D Fact). Both proxy-based variants produce significantly blurrier results, although the factorized version performs slightly better.}
\vspace{-0.1in}
\label{fig:ablation_shell}
\end{figure}

\paragraph{Appearance Paramatrization}
We explore an alternative appearance parametrization of Eq.~\ref{eq:radiance_recovery} by constraining $\xx$ to be on a spherical shell tightly enclosing the luminaire. Fig.~\ref{fig:ablation_shell} shows appearance comparisons of the \textsc{Rough Flower} and \textsc{Modern} luminaires using two shell-based variants. The first variant \textsc{Shell 4D} learns the four-dimensional PDF using a single NF architecture, resulting in coarse estimates. The second variant \textsc{Shell 4D Fact} follows Eq.~\ref{eq:factorization} to factorize the objective as marginal and conditional 2D distributions, each modeled with a normalizing flow. This representation allows the conditional NF to encode directions using a learnable 2D dense grid encoding over sphere shell positions, improving the estimated luminaire. Our proposed formulation further improves on this estimate by learning the target directly on the mesh surface, which eliminates the encoding quality degradation as the predicted surface gets further away from the enclosing shell.

\begin{figure}
\centering
\includegraphics[width=\linewidth]{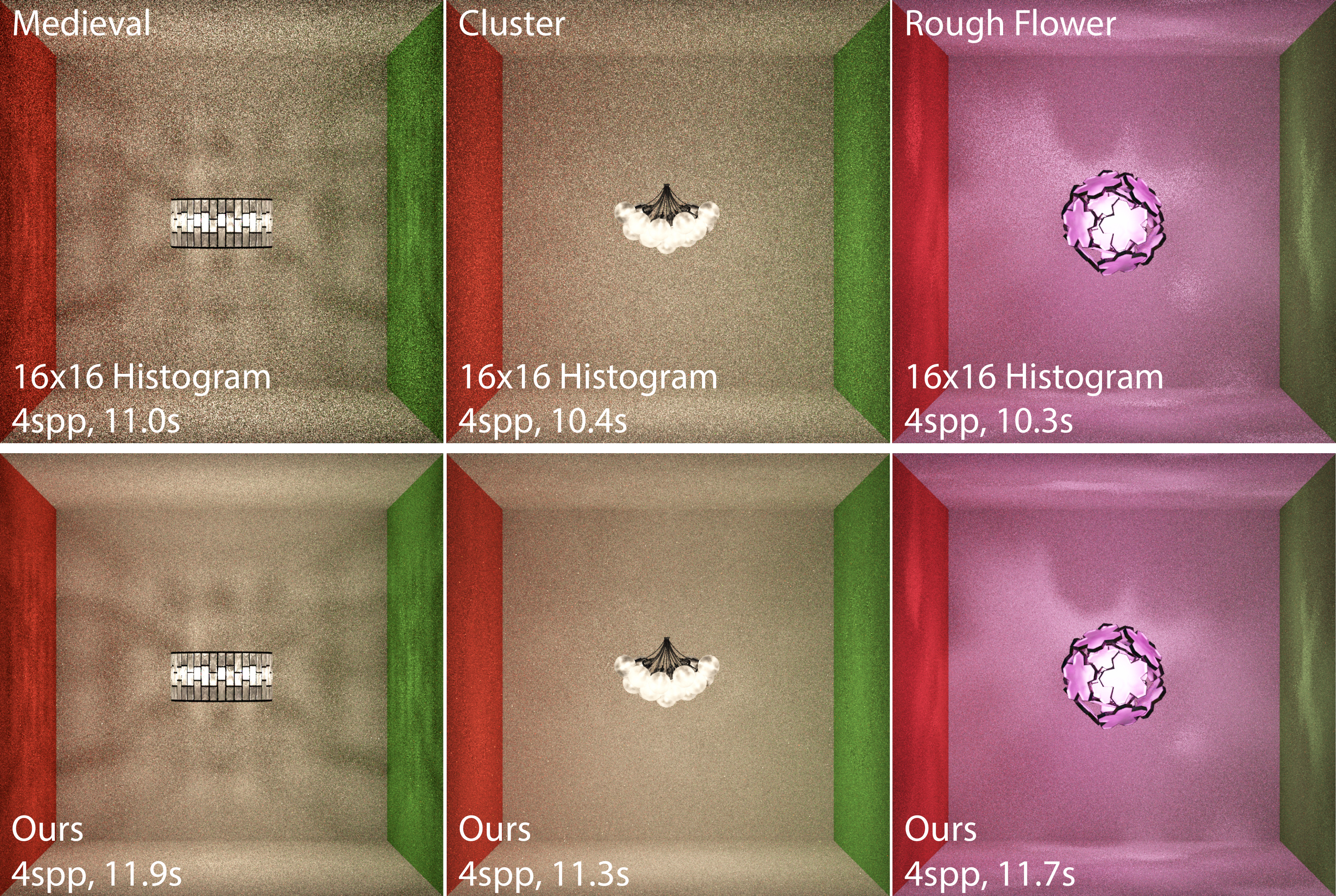}
\vspace{-0.25in}
\caption{\rev{Comparison of importance sampling using our normalizing flow against a variant using $16 \times 16$ histograms. Our method produces results with less noise because it produces higher quality distributions with a small runtime overhead.}}
\label{fig:ablation_isarch}
\end{figure}

\paragraph{Importance Sampling Architecture}
We analyze the choice of architecture for our importance sampling (IS) network in Fig.~\ref{fig:ablation_isarch}. We compare our discretization-free normalizing flow against a discrete $16 \times 16$ histogram-based importance sampling model. Note that, this histogram baseline is used only for sampling and is different from our fast rendering variant which uses a discretized representation for both sampling and radiance evaluation. As shown, our results with the normalizing flow are generally less noisy, since the estimated distributions are of higher quality.

\paragraph{\rev{Bias Analysis of Ours Fast}}

\begin{figure}
\centering
\includegraphics[width=\linewidth]{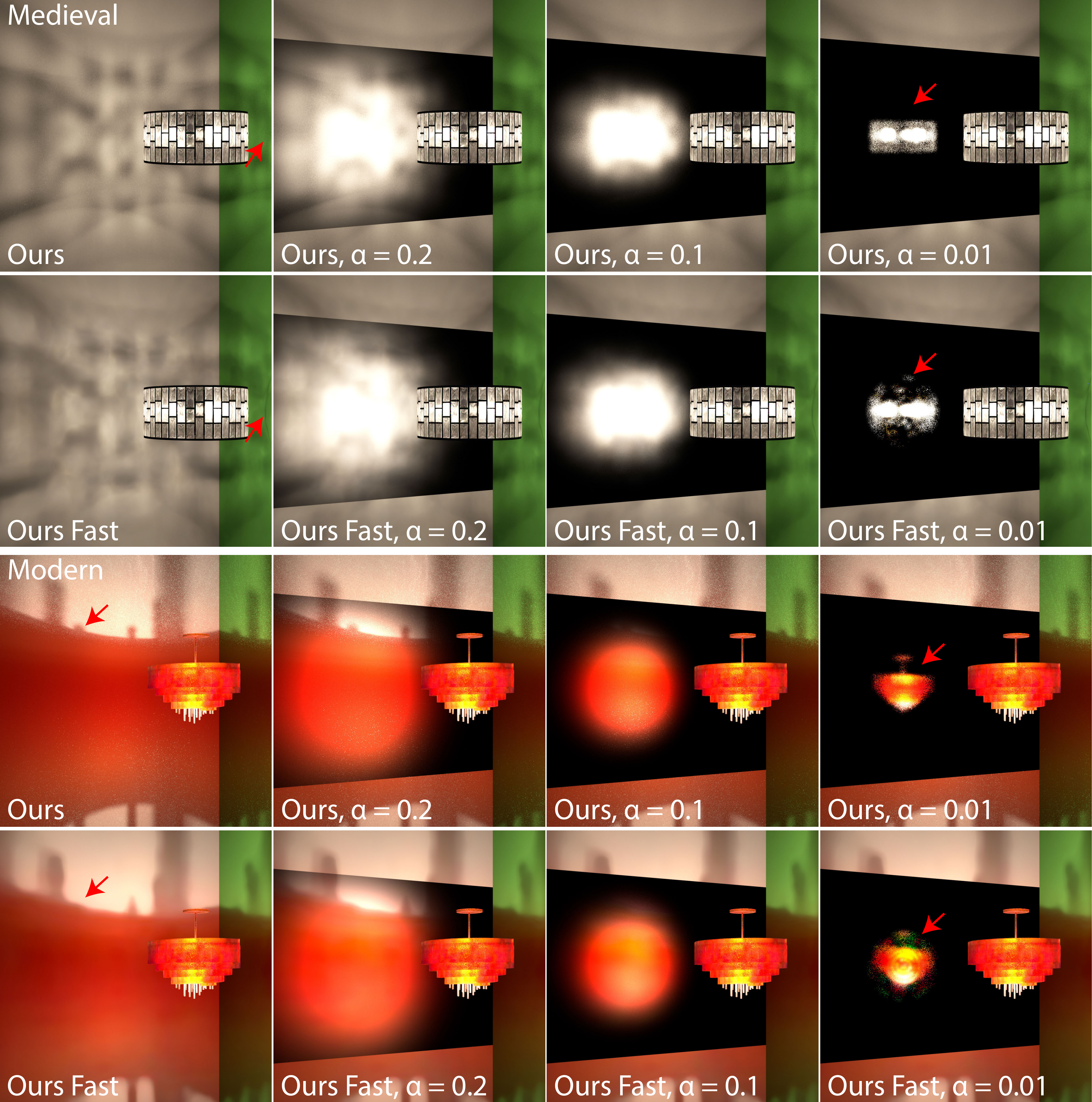}
\vspace{-0.25in}
\caption{\rev{Equal-sample comparison of first-bounce direct illumination produced by the two variants of our method, Ours and Ours Fast. Ours produces consistent illumination even on low-roughness surfaces because it traces an additional ray to intersect the luminaire and evaluates the appearance network to obtain radiance. Ours Fast produces plausible estimates on diffuse surfaces and reflective surfaces with higher $\alpha$, but discretization artifacts become more noticeable at lower roughness values. For both variants, direct appearance along camera rays is evaluated using the appearance and transparency networks.}}
\label{fig:ablation_variant}
\end{figure}

\rev{We compare the illumination behavior of both variants of our method (\textit{Ours} and \textit{Ours Fast}) at equal sample counts in Fig.~\ref{fig:ablation_variant}. We render only first-bounce direct illumination to highlight the differences during next-event estimation. The first column shows renders of the \textsc{Medieval} and \textsc{Modern} luminaires in the cornell box. Starting from the second column, we vary the roughness of a reflective surface placed next to the luminaire to analyze the effects of the discretization used by our \textit{fast} variant. Our method produces consistently accurate illumination across surfaces of varying roughness because it validates sampled directions by intersecting them with the luminaire and evaluates radiance using the appearance network. Our fast variant produces less noisy renderings, but it may sample directions outside the luminaire surface and use its low-resolution radiance estimates. This results in color artifacts that are most noticeable in specular reflections.}

\subsection{Limitations and Future Work}

Despite its advantages, our method has several limitations. 
\rev{First, our importance sampling strategy assumes that shading points lie outside the luminaire's bounding sphere, limiting support for wall-mounted luminaires and luminaires placed near surrounding geometry. Second, our implementation targets modern GPU-based rendering pipelines, in which neural network inference is practical, making it less suitable for CPU-based pipelines. Third, as with other neural approaches~\cite{Zhu_2021_ToG, Condor2022}, our method requires longer pre-computation than some non-neural alternatives~\cite{velazquez2015complex, WCL}, potentially slowing design iterations. Finally, as with previous pre-computation techniques~\cite{WCL, Zhu_2021_ToG, Condor2022, velazquez2015complex}, our method does not account for scene illumination reflected by the luminaire surfaces. Modeling this interaction would require a full 8D light-transport function relating incoming and outgoing radiance over luminaire surfaces, as demonstrated by 8DNA~\cite{Wu2026_8DNA} for neural assets, which we leave for future work.}


\section{Conclusion}

In this paper, we introduced PureLight, a novel neural representation for complex luminaires that overcomes the limitations of both traditional Monte Carlo rendering and previous neural solutions. By framing the problem as a distribution learning task, we successfully extract the appearance of arbitrarily complex light sources, including ones that are virtually impossible to render using standard Monte Carlo estimators. Our approach exclusively leverages light tracing (which is always available and efficient) to build an accurate neural model of exitant radiance as a product of a learned normalizing flow and a flux term.

We further introduced a distilled MLP for fast radiance evaluation; for efficient direct illumination queries, we introduced an importance sampling network and a fast discretized representation, significantly reducing variance and enabling noise-free direct illumination even at low sample counts. These components, with an additional transparency network, ensure that complex, physically-accurate luminaires can be integrated into standard rendering pipelines without prohibitive computational costs and high variance.

Our method bridges the gap between the design of detailed 3D light assets as precise CAD models with highly specular materials, and their practical use in efficient rendering systems. By shifting the complexity from real-time path estimation to a precomputed neural representation, we enable artists and engineers to incorporate arbitrarily sophisticated luminaire designs, such as tiny LEDs and complex glass housings. We believe our distribution-based learning strategy offers a scalable path forward for representing a wide variety of complex luminaires in the field of physically-based rendering.

\begin{acks}
This project was funded in part by the NSF CAREER Award $\#2238193$. Portions of this research were conducted with the advanced computing resources provided by Texas A\&M High Performance Research Computing. We are grateful to Zhu et al.~\shortcite{Zhu_2021_ToG} for releasing their source code and providing the \textsc{Elevator Room} and teaser scenes used in our experiments. We would like to thank the following artists for sharing scenes and models that appear in our figures: Jay-Artist (\textsc{Kitchen}), SlykDrako (\textsc{Bedroom}), Wig42 (\textsc{Dining Room} and \textsc{Horse Room}), and Benedikt Bitterli (\textsc{Cornell Box}).
\end{acks}

\bibliographystyle{ACM-Reference-Format}
\bibliography{bibliography}

\end{document}